 \documentclass[final,5p,times,twocolumn]{elsarticle}

\usepackage{lineno,hyperref}
\modulolinenumbers[5]

\journal{MDPI}

\usepackage{numcompress}

\usepackage{xcolor}

\usepackage{epsfig}

\usepackage{amssymb}

\usepackage{float}

\begin{document}

\begin{frontmatter}



\title{A hard open X-band RF accelerating structure made by two halves}

\author{Bruno Spataro $^{a}$, Mostafa Behtouei $^{a}$, Fabio Cardelli $^{a}$, Martina Carillo $^{b}$, Valery Dolgashev $^{c}$, Luigi Faillace$^{a}$, Mauro Migliorati $^{b}$, and Luigi Palumbo$^{b}$\\\vspace{6pt}}

\address{$^{a}${INFN, Laboratori Nazionali di Frascati, P.O. Box 13, I-00044 Frascati, Italy};\\
 $^{b}${Dipartimento di Scienze di Base e Applicate per l'Ingegneria (SBAI), Sapienza University of Rome, Rome, Italy};\\
 $^{c}${SLAC National Accelerator Laboratory  };\\
  }

\begin{abstract}
High-gradient linacs of next generation require novel accelerating structures which are compact, robust and cost-effective. Dedicated research and development have been launched in the linear-collider community. This paper focuses on the technological developments directed to show the viability of novel welding techniques and related applications, in order to benefit from the superior high-gradient performance of accelerating structures made of hard-copper alloys. The  structure geometry that we propose allows getting a high longitudinal shunt impedance of the accelerating mode and increasing the mode separation frequencies.
\end{abstract}


\end{frontmatter}

\section{Activities on X band structures made of two halves}
High-gradient linacs are sought for various applications, from high-energy physics, industry and medicine, and require innovative accelerating structures. An intense activity on  research and development has been launched in the linear-collider community, as in \cite{ref1,ref2,ref3,ref4,ref5,ref6}.  A continuous collaboration on the study of various geometries, materials, surface processing techniques and technological developments of accelerating structures has involved, for more than a decade, the SLAC National Accelerator Laboratory in the USA, the Italian Institute of Nuclear Physics - National Laboratories of Frascati (INFN-LNF), and the High Energy Accelerator Research Organization (KEK) in Japan \cite{ref2}. 

\begin{figure*}[t]
\begin{center}
\includegraphics[width=6cm]{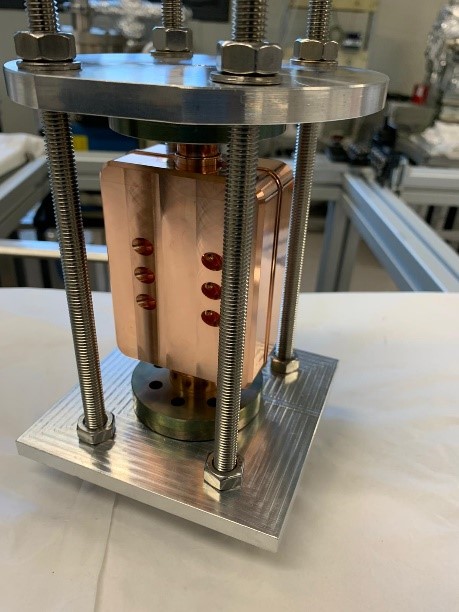}
\includegraphics[width=6cm]{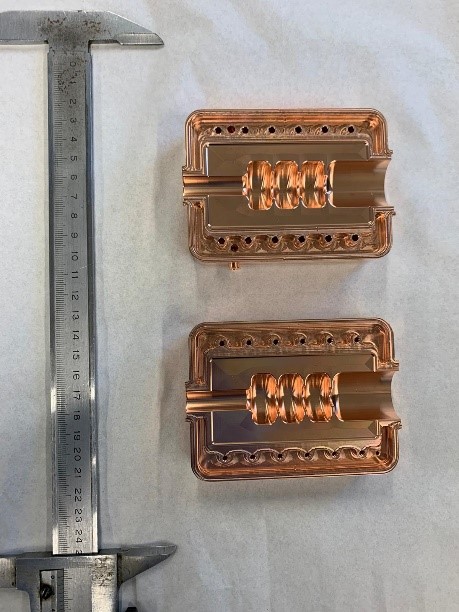}\\
\includegraphics[width=12cm]{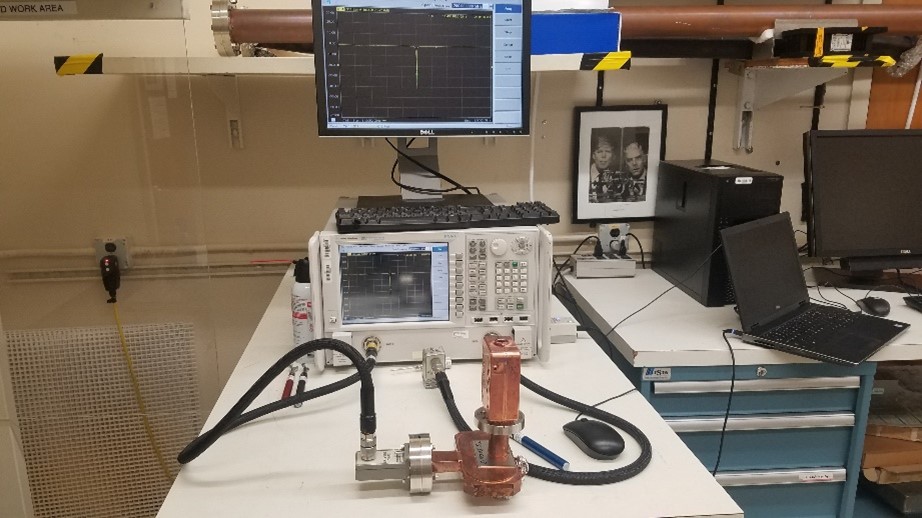}
\caption{Picture of the two halves structure under test.}
\label{fig:Figure1}
\end{center}
   \end{figure*}

This paper is part of this study and it focuses on the technological developments directed to show the viability of novel welding techniques \cite{ref3,ref5,ref6}, and related applications, in order to benefit from the superior high-gradient performance of accelerating structures made of hard-copper alloys. The technological activity of testing high-gradient RF sections, in particular at 11.424 GHz, began at INFN-LNF in the framework of the above-mentioned collaboration, and it is related to the investigation of breakdown mechanisms, which limit the high gradient performance of these structures. It consists in the design, construction and high-power experimental tests of standing wave (SW) 11.424 GHz (X-band) accelerating cavities with different materials and methods. 
Figure \ref{fig:Figure1} shows the two halves of the X-band assembled structure under RF low level test.

The goal of the collaboration is to assess the maximum sustainable gradients with extremely low probability of RF breakdown in normal-conducting high-gradient RF cavities. The most common bonding techniques, used worldwide, are high-temperature brazing and diffusion bonding. The brazing and the diffusion bonding are performed inside a high-temperature furnace. On the other hand, experimental results with hard copper cavities, conducted at SLAC, CERN and KEK \cite{ref1,ref2} have shown that hard materials sustain higher accelerating gradients for the same breakdown rate. Therefore it would be better to avoid high temperature processing of the cavities to benefit from superior high gradient performance of hard copper alloys. 

In this framework, we are conducting experiments that involve the Electron Beam Welding (EBW) and Tungsten Inert Gas (TIG) processes, which allow us to build practical, multi-cell structures made with hard copper alloys, in order to increase their RF performance against the soft ones. For this purpose, open structures made of two halves have been investigated and fabricated. Details on the fabrication procedure of “split” X Band structures made of two halves, or \textquotedblleft{open}\textquotedblright structures, by using the TIG method are given in ref \cite{ref7}. 

In this paper, we present the RF characterization and low-power RF tests of a TIG welded two-halves split hard-copper structure \cite{Zha1,Zha2} that will be consequently employed for high-gradient tests and for the study of the RF breakdown physics. To this aim, the structure geometry that we propose (and shown in Fig. 2 -> cavity design) allows getting a high longitudinal shunt impedance of the accelerating mode, increasing the mode separation frequencies, and improving the operation vacuum level. 
In addition, intense beam currents and multibunch operation are essential features, for example for increasing the luminosity of a linear collider, but beam current wakefields and coupled-bunch mode instabilities, which mostly arise from the parasitic modes of the accelerating structures, can put a limit to the accelerator performance. Hence our main interest is also to detune the cavity in order to reduce the beam instability by using a novel simpler technique and dedicated absorbers of the higher order modes (HOMs). The two cavity halves are aligned and clamped together by means of male-female matching surface, and the clamping is obtained with stainless screws and TIG welded at COMEB \cite{ref8} on the outer surface. Preliminary low-power RF measurements are in agreement with the simulated ones. The estimation of mode separation is described later in this paper. The activity to fabricate a section with 4 quadrants is in progress, too.

\section{Low level RF measurements}

It consists of three cells: the middle one is the high gradient cell, while the first and third ones are end-cells. The peak on-axis electric field in the middle cell has to be two times higher than the end-cells.

\begin{figure*}[t]
\begin{center}
\includegraphics[width=6cm]{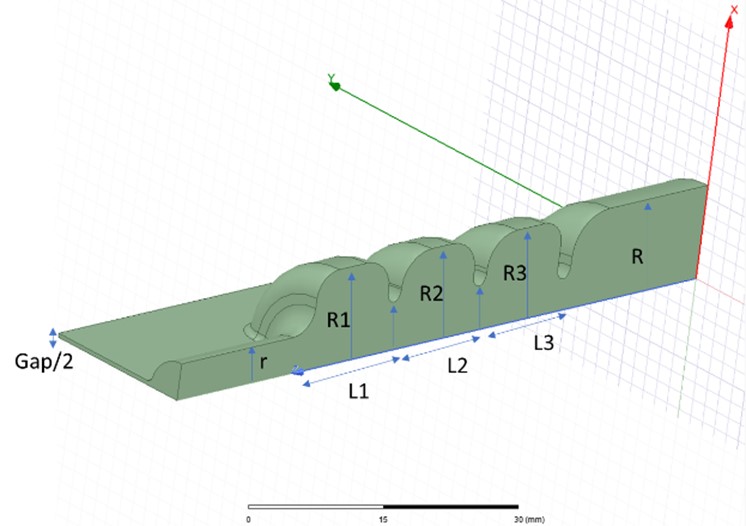}
\includegraphics[width=4cm]{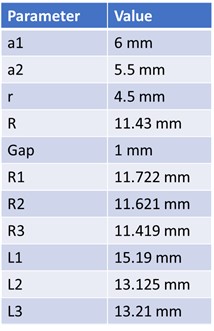}
\caption{one-quarter of two half structure sketch for the simulation studies by using ANSYS HFSS software. }
\label{fig:Figure2}

\includegraphics[width=6cm]{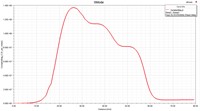}
\includegraphics[width=6cm]{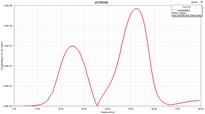}
\begin{center}
\includegraphics[width=8cm]{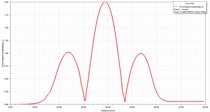}
\end{center}
\caption{Longitudinal field profile of the structure modes: a) $0$ mode , b) $\pi/2$ mode,  c) $\pi$ mode}
\label{fig:Figure3}
\end{center}
   \end{figure*}

Figure \ref{fig:Figure2} reports one-quarter of two half structure sketch for the simulation studies performed with ANSYS HFSS software \cite{ref9}.

\begin{figure*}[t]
 \begin{center}
\includegraphics[width=13cm]{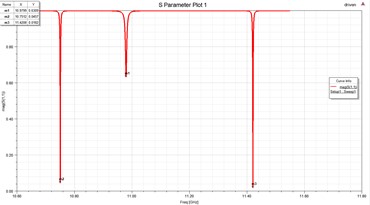}
\caption{Simulated structure frequencies spectrum}
\label{fig:Figure4}
 
\includegraphics[width=7cm]{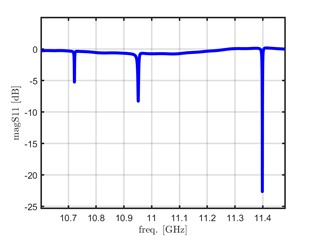}
\includegraphics[width=6cm]{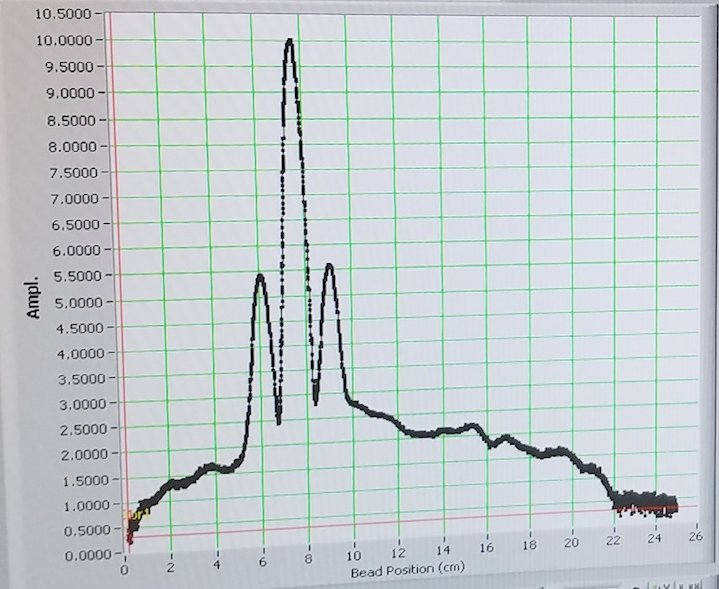}
\caption{Measured frequency spectrum and longitudinal field profile of the $\pi$-mode}
\label{fig:Figure5}
\end{center}
   \end{figure*}

Figures \ref{fig:Figure3}a, \ref{fig:Figure3}b and \ref{fig:Figure3}c show the longitudinal electric field profile of the $0$, $\pi/2$ and $\pi$ modes.
The simulated frequencies spectrum is reported in Figure \ref{fig:Figure4}.

Figure \ref{fig:Figure5} shows the measured frequencies spectrum and longitudinal field profile of the $\pi$-mode obtained with the bead pull technique by using two antennas.

The numerical estimations of mode frequencies and the longitudinal field profile of the $\pi$-mode are in good agreement with the experimental ones.

In Table \ref{tab:Comparison} we report frequencies and quality factors of the resonant modes, obtained from simulations, for the split open structure in comparison with the same cavity designed with the conventional “closed” approach.  

\begin{table}[h]
{\scriptsize \begin{tabular}{ccccc}
\hline
Mode &  F [MHz]&F [MHz]& Q*&Q*
\\
&(two halves)&(closed structure)&(two halves)&(closed structure)
\\
\hline
$0$&10,749&10,760&10,520&10,615\\
$\pi/2$&10,979&10,984&10,213&10,306\\
$\pi$&11,418&11,420&10,512&10,610\\
\hline \hline
\end{tabular}
*Quality factor }
\caption{
 Comparison between mode frequencies and quality factors of the closed structure and two halves one.}
\label{tab:Comparison}
\end{table}

The frequency separation between the $\pi$ - $\pi/2$ and $\pi/2$ – $0$ modes is about 440 MHz and 230 MHz, respectively, in the case of  the two halves structure, while it is equal to about 435 MHz  and 225 MHz for the closed one. The two-halves structure cell-to-cell coupling coefficient is estimated to be 6.1\% , while the one for the other structure is 6 \%.  It is important to notice that the amount of frequency separation of each mode is proportional to the cavity form factor Rsh/Q \cite{ref10} which depends on the cell-to-cell coupling coefficient. In addition, from a comparison between the  two halves structure and the closed one, we also observe that the frequency separation is wider in the case of the two-halves structure since  the capacitive effect is stronger with respect to that of the closed cavity.  In particular, the 0-mode of the open cavity is the most affected with a deviation which is 11 MHz greater with respect to the homologous separation in the closed one. By increasing the number of splittings, we are able to increase the mode frequency separation accordingly, too.  

As additional information, the frequency separation for this structure results to be 25\% larger than the one achieved in other standard brazed structures [10] with a rough hard edge geometry   but without splitting~\cite{ref11}.  The measured quality factor of the $\pi$ mode of the two-halves structure, obtained by using two antennas, is about Q $\sim$ 6900, which is 30 \% lower with respect to the numerical predictions due to the bad relatively poor electrical contact during low-power measurements. Our next step is to repeat the RF characterizations after having welded the two halves and also by using the input power mode launcher.  

We have also scheduled to construct another structure made of 4 quadrants, in order to confirm the expected wider amount of mode frequency separation. Moreover, for reasons related to both the feasibility of the hard structure and a high shunt impedance, we believe that we need to find a compromise between the  of  detuning effects and the number of the structure subdivisions. This approach can be extended to the cavity resonators used in circular accelerators, in order to eliminate or strongly reduce the longitudinal coupled-bunch instability associated with the accelerating mode.

\section{Possible applications of split open structures}
The next generation of accelerators is highly demanding in terms of maximizing accelerating gradients, minimizing overall machine length and cost, improving the beam quality, reducing beam loading effects, and so on. In case of circular accelerators, one important mechanism which produces beam instability and power losses consists in the possibility of the charged particle to excite high order resonant fields in RF accelerating cavities when passing through them. This is possible only for those accelerating  structures having resonant modes in frequency ranges which overlap beam Fourier spectrum. If the beam loading on a large electron /positrons storage ring is strong, the accelerating mode of a normal conducting cavity gives rise to a coupled-bunch instability. 

Additionally, in  recent years, we have seen a revolution in high brightness electron beams because of the maturation of RF photo-injector performance. In order to improve the electron beam brightness,  larger mode separation is also needed.  Also in pre-buncher and chopper cavities of linear accelerator, operating at high currents, an RF detuning is required in order to reduce the interaction with the higher order modes which affect the beam quality.  These problems can be cured by using the split open structure approach which gives also the possibility of improving the vacuum level of at least a factor ten with respect the usual solution by inserting an additional vacuum chamber connected to the beam pipe. For a fixed geometry of the structure (which means a fixed form factor  ratio Rsh/Q ) and for a given beam current and cavity peak voltage, the frequency detuning effect can be determined, for example, as shown in \cite{ref10}. Intense investigations on the absorbers for damping  higher order modes, are in progress, too.


\section*{Acknowledgment}

The authors would like to thank James Rosenzweig from UCLA for his useful discussion and helpful advice.

\end{document}